\documentclass[a4paper,11pt]{article}
\pdfoutput=1 

\usepackage{jheppub} 
\usepackage[T1]{fontenc} 

\usepackage{lineno}
\usepackage{bbm}

\usepackage{orcidlink}
\newcommand{\orcid}[1]{\orcidlink{#1}}

\usepackage[most]{tcolorbox}
\usepackage{xcolor}
\usepackage{enumitem}

\usepackage{tikz}
\usetikzlibrary{arrows.meta, positioning, fit, calc}
\usepackage{algorithm}
\usepackage{algpseudocode}
\usepackage[most]{tcolorbox}
\usepackage{comment}
\usepackage[english]{babel}
\usepackage{color}
\usepackage{xcolor}
\usepackage{adjustbox}
\usepackage{float}
\usepackage{subcaption}
\usepackage{booktabs}
\usepackage{multirow}
\usepackage{makecell}
\usepackage{xcolor}
\usepackage{colortbl}
\usepackage{amsmath}
\usepackage{amssymb}
\usepackage{array}

\usetikzlibrary{shapes.geometric}

\usepackage{listings}
\usepackage{textcomp}

\lstdefinestyle{maverickprompt}{
  basicstyle=\ttfamily\footnotesize,
  breaklines=true,
  breakatwhitespace=true,
  columns=fullflexible,
  keepspaces=true,
  showstringspaces=false,
  upquote=true,
  literate=
    {°}{{$^\circ$}}1
    {±}{{$\pm$}}1
    {γ}{{$\gamma$}}1
    {θ}{{$\theta$}}1
    {μ}{{$\mu$}}1
    {–}{{--}}1
    {—}{{---}}1
    {≈}{{$\approx$}}1
}

\title{Towards Retrieval Augmented Generation in High-Energy and Astroparticle Physics}

\author[a]{Jacky Kumar \orcid{0000-0001-9053-0731},}
\author[b]{Sajan Kumar \orcid{0000-0002-5167-1221}}

\affiliation[a]{Physique des Particules, Universite de Montreal, Montreal, QC, Canada}
\affiliation[b]{Department of Physics, University of Maryland, College Park, Maryland, USA}

\note{\emph{Author list is in alphabetical order. Both authors contributed equally to this work.}}

\abstract{
The high-energy and Astroparticle physics literature has grown into an enormous body of knowledge. Gaining a comprehensive understanding of this literature is an important step in research, but is a challenging task. Before making meaningful advances, it is important to establish what has already been done and identify open questions. However, this process is becoming increasingly difficult given the sheer volume of publications. Conducting a focused literature survey on a narrow topic is particularly challenging. We develop an open-source Retrieval Augmented Generation (RAG) pipeline to produce concise, targeted summary reports with citations, grounded in the database of arXiv papers, enabling researchers, editors, and referees to rapidly orient themselves within any corner of the field. Specifically, we embed approximately 230K hep-ph and astro-ph.HE papers to enable a hybrid retrieval that captures both keyword matches and semantic similarity. The fused candidate set is then refined using a cross-encoder reranker. Following retrieval, the papers are passed to a Large Language Model (LLM), for which we use Google DeepMind's open-source \textit{Gemma-4-E4B} model. The LLM first acts as a judge to evaluate the relevance and then synthesizes a coherent and grounded report with references. This modern AI-driven approach allows us to go beyond the capabilities of classical keyword search on arXiv or Inspire-HEP.
}


\begin{document}
\preprint{UdeM-GPP-TH-26-314}
\maketitle
\newpage

\section{Introduction}

Easy access to scientific knowledge is the critical foundation for 
its dissemination. But for fields like high-energy physics/astroparticle physics (HEP \& Astro) this 
is becoming very challenging due to the rapid growth of the literature. Throughout this work, we use HEP-Astro to refer collectively to high-energy physics and astroparticle physics.
In 2025, the arXiv database hosted more than 2.6M research papers 
across various disciplines. 
In order to find 
a paper on a narrow topic, researchers mainly rely on a classical 
keyword search from databases such as arXiv, Inspire-HEP, or Google Scholar 
to locate articles relevant to a narrow topic of interest.  However, these approaches are 
limited as they do not capture the semantic meaning.

LLMs such as Claude, GPT, Gemini, and Llama have
significantly changed how we access, interpret, and generate knowledge. Despite
being very broad, their parametric knowledge is fixed at training
time, so they cannot know about research articles published afterward. When asked about
material outside of this knowledge, LLMs can respond with incomplete information or
produce factually incorrect responses.
Continuously updating LLMs on newly published papers is impractical, given that they appear everyday.
Therefore, some commercial LLMs partially resolve this by using web search tools at
inference time, but general web search is not tailored to specific type of  
literature. In addition,
commercial models require subscription fees or API costs, which can be a barrier
to building domain-specific applications. Fine-tuning LLMs on domain
corpora~\cite{Richmond:2025lzg, nguyen-etal-2023-astrollama} addresses part of
this problem, but it is computationally expensive and is
hard to keep up to date.

RAG offers a powerful alternative which
couples the generative capabilities of LLMs with a dynamic 
retrieval mechanism that fetches semantically relevant documents from an external source at inference time \cite{rag, gao2024retrievalaugmentedgenerationlargelanguage} (see also Ref.~\cite{ragreview} for review). The retrieved passages are then injected into the LLM context window, grounding it's generated output in verifiable evidence which could be domain specific.

Indeed, the importance of RAG has been recognized by the {particle and astro physics community, where it has been utilized for the knowledge retrieval from domain-specific
 databases at several fronts}.  For instance, the large scale experiments have used RAG for navigating 
information from internal documentation, technical reports, and answer context aware queries, see for example DUNE-GPT \cite{rafique2026largelanguagemodelintegration} for DUNE experiment,  MITRA \cite{mallampalli2026mitraaiassistantknowledge} for CMS, and chatlas \cite{DalSanto:2935252} dedicated for ATLAS experiments at the LHC.{Similar approaches have also been explored in astronomy, such as Pathfinder \cite{Iyer_2024}, a semantic framework designed to support literature review and knowledge discovery in the astronomical literature.}

In addition to these RAG-based applications, significant effort has also been devoted on foundational aspects upon which effective retrieval depends, in specific domains. This includes new text embedding models such as PhysBERT~\cite{hellert2024physberttextembeddingmodel} and Astro-HEP-BERT~\cite{simons2024astrohepbertbidirectionallanguagemodel} pretrained on large corpora of physics literature and designed to capture the semantic nuances of HEP-Astro specific terminology. Evaluation pipelines have also been developed to assess the quality of retrieval in scientific question answer settings~\cite{Xu:2025uwt}. In this regard,  Ref.~\cite{Jiang:2026pkh} introduces an agentic hybrid RAG framework for muon collider research alongside the dedicated benchmark for retrieval-augmented question answering .

Since rapidly evolving HEP-Astro literature is posing significant challenges 
to both experienced and new researchers to stay up to date. 
An advanced AI-driven method that can efficiently retrieve and synthesize knowledge 
from arXiv HEP-Astro database is therefore a valuable and timely step. To best of our 
knowledge, until now, not much attention has been given to develop a RAG method
for HEP-Astro subdomains hep-ph and astro-ph.HE. 

In this work, we develop a RAG pipeline dedicated to these fields. 
{Given a user's scientific query, using our pipeline}, relevant papers are semantically retrieved from arXiv papers embedded using \textit{Specter2} model \cite{specter2}.
These papers are then provided as context to an open-source LLM, such as \textit{Gemma4} \cite{gemma4} or \textit{Llama} \cite{llama},  to synthesize a concise summary report including citations. 
There can be multiple applications of this pipeline. 
For example, it can help researchers get insights from the literature, perform gap analysis 
or perhaps even identify novel  directions by knowing what has been already done. 
It can also facilitate the peer review process. We stress that our goal is not to replace expert judgment, but to reduce the {effort} for literature search and provide evidence based assistance to the researchers under a rapidly growing publications under the influence of AI.

This article is organized as follows, Sec.~\ref{sec:method} is devoted to methodology,  where we detail 
the key ingredients of our RAG pipeline such as data preparation, 
creating embeddings using \textit{Specter2}, storing them in a vector database. Sec.~\ref{sec:retrieval} details the retrieval process. In Sec.~\ref{sec:bench-eval}, we give details of benchmark and evaluation of RAG pipeline. In Sec.~\ref{sec:generation}, we discuss the report synthesis with an LLM. 
Finally, we conclude and briefly discuss possible future improvements in Sec.~\ref{sec:conclude}.

\section{Dense Embeddings of HEP-Astro Literature}
\label{sec:method}
In this section, we discuss the different components of our RAG pipeline. For a quick overview of the embedding and retrieval procedures, we refer to Fig.~\ref{fig:embeddings} and Fig.~\ref{fig:retrieval}, respectively. In the next subsection, we first discuss our database of arXiv HEP-Astro articles, which serves as the literature input for the embedding procedure.

\subsection{HEP-Astro literature}
Our ultimate goal is to be able to synthesize short review report on a given HEP-Astro topic 
for the purpose of gap and novelty analysis. Such reports can also be useful for the 
peer-review process and to find new research directions. Therefore, it is very important 
that the report be grounded in published work and covers important previous contributions, 
with appropriate references.  

For this purpose, we believe that the full text of papers is not necessary, but, only the title, abstract, and associated metadata are sufficient. Indeed, these are good representation of the paper and usually contain the overall motivation and new findings.

We construct our input from a Kaggle dataset, which contains all the required features mentioned above \cite{arxiv_org_submitters_2024}. This dataset is distributed in {\tt json} format and updated weekly.  From this dataset, we filter papers belonging to the subdomains hep-ph and astro-ph.HE, which align with our domain expertise and include the preprints submitted to arXiv until the end of 2025. After filtering, our HEP and Astro knowledge base consists of total 253,606 articles. Of these, we there are {141,160} hep-ph and {45,045}  astro-ph.HE primary category papers and {52,762} hep-ph and {22,256}  astro-ph.HE secondary category papers. Note that these numbers are not supposed to add up to 253,606, because a paper can be hep-ph and astro-ph.HE secondary at the same time. Indeed, we checked that there are 186,305 only hep-ph related, 59,684 only astro-ph.HE related papers, and 7617 papers contain both hep-ph and astro-ph.HE -- which add up to 253,606.
\begin{figure}[t] 
  \centering
    \includegraphics[width=0.8\linewidth,height=\textheight,keepaspectratio]{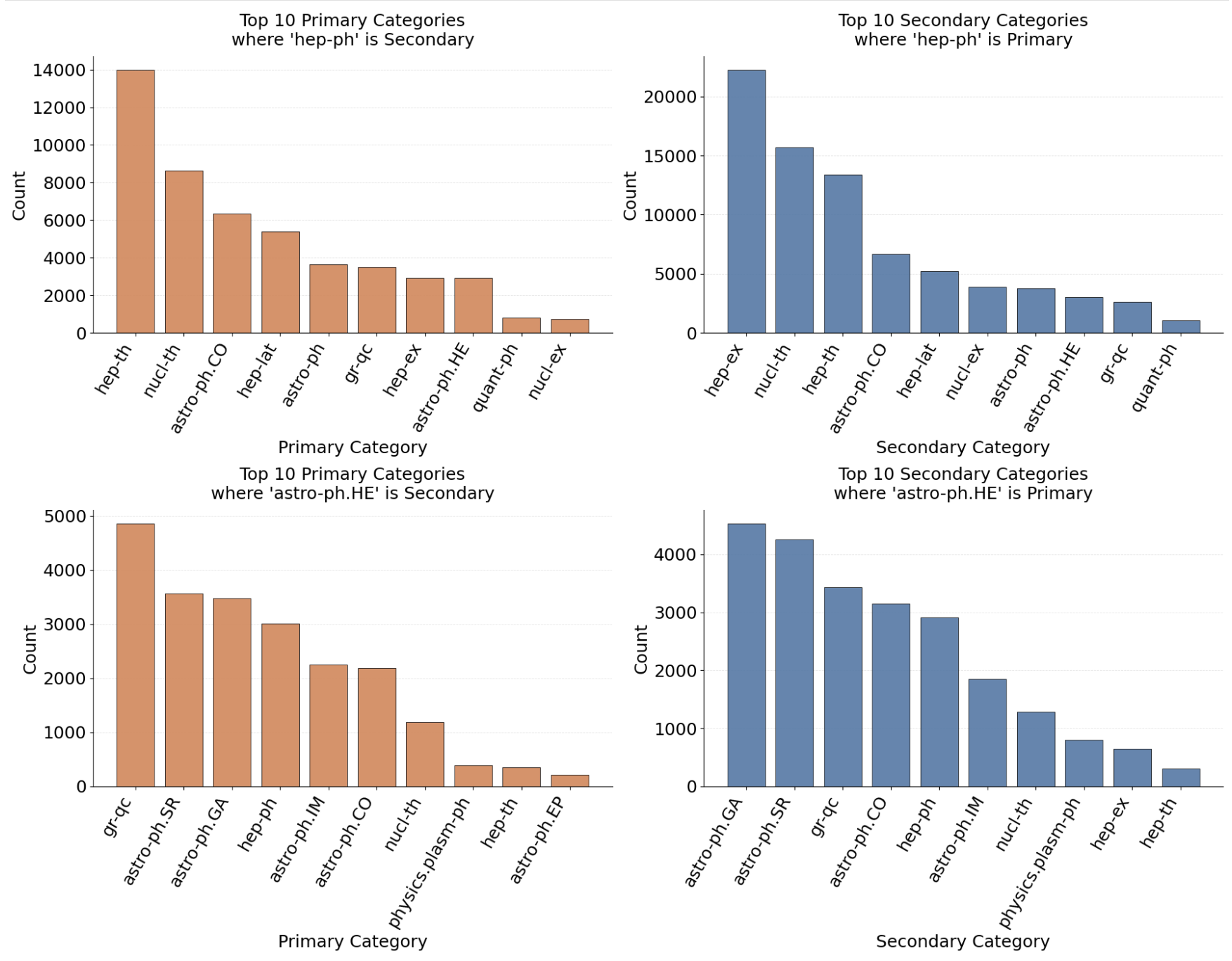}
  \caption{\small Top-10 secondary (primary) categories when hep-ph or astro-ph.HE are primary (secondary) category papers in our database.  }
 \label{fig:dataset}
\end{figure}
In Fig.~\ref{fig:dataset}, we show a barcharts for the top-10 secondary categories in hep-ph primary category papers, as well as the top-10 primary categories when hep-ph is secondary. A similar barcharts for astro-ph.HE are also shown.

\subsection{\textit{Specter2}, \textit{Qwen3}, and \textit{Nomic} embeddings} 

To create embeddings for the HEP-Astro dataset, we use the \textit{Specter2} embedding model, which is an open-source 
encoder model \cite{specter2}. Its pretrained weights are publicly available at Hugging Face\footnote{\url{https://huggingface.co/allenai/specter2}}. \textit{Specter2} is trained on title and abstract of scientific papers in various fields including physics.
Therefore, the resulting embeddings are more suitable for scientific document retrieval as compared to general purpose sentence encoder models. 

\textit{Specter2} model's pretraining objective is the citation relationships; in other words, papers that cite each other or are related tend to be closer in the embedding space, which is 768 dimensional.  
Following pretraining on 6M citation triplets, it is also fine-tuned using task-specific adapters trained on the SciRepEval benchmark \cite{Singh2022SciRepEvalAM}.
We have added so-called the \textit{proximity} adapter on the top of \textit{Specter2} base model to embed abstract and titles and \textit{adhoc query} adapter to embed the query.   

We will also compare the performance of \textit{Specter2} with two general purpose embedders \textit{Qwen3} \cite{qwen} and \textit{Nomic} \cite{nomic}. \textit{Qwen3} -- a state-of-the-art embedding, is representative of large, general purpose LLM derived embedders and is based on \textit{Qwen3} foundational models. \textit{Qwen3} spans model size of 0.6B, 4B and 8B parameters, but we use \textit{Qwen3-0.6B}. 
On the other hand, \textit{Nomic} is English text embedding supporting long context length 
having only 137M parameters.  

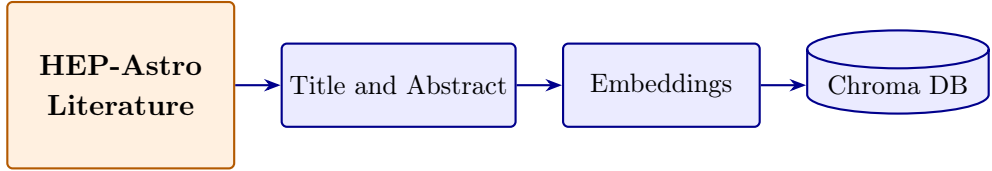
\begin{figure}[H]
\centering
\begin{tikzpicture}[
    font=\small,
    node distance=6mm,
    kb/.style={
        draw=orange!70!black, thick, rectangle, rounded corners=2pt,
        fill=orange!12, minimum width=3.0cm, minimum height=2.2cm,
        align=center, font=\bfseries, inner sep=4pt
    },
    box/.style={
        draw=blue!55!black, thick, rectangle, rounded corners=2pt,
        fill=blue!8, minimum width=2.6cm, minimum height=1.1cm,
        align=center, inner sep=3pt
    },
    db/.style={
        draw=blue!55!black, thick, cylinder, shape border rotate=90,
        fill=blue!8, aspect=0.25, minimum width=2.4cm, minimum height=1.1cm,
        align=center, inner sep=2pt
    },
    arrow/.style={-{Stealth[length=2.2mm]}, thick, blue!55!black}
]

\node[kb] (kbdoc) {HEP-Astro\\ Literature};
\node[box, right=of kbdoc] (text) {Title and Abstract};
\node[box, right=of text] (embed) {Embeddings};
\node[db, right=of embed] (chroma) {Chroma DB};

\draw[arrow] (kbdoc) -- (text);
\draw[arrow] (text) -- (embed);
\draw[arrow] (embed) -- (chroma);

\end{tikzpicture}
\caption{\small Procedure for creating embeddings of dataset containing title and abstract of HEP and Astro papers and their storage in the chroma vector database.}
 \label{fig:embeddings}
\end{figure}

Each paper is represented by a single input sequence constructed by concatenating its title and abstract. The resulting text is embedded and stored in the ChromaDB vector database. Metadata fields are stored separately alongside each embedding and are used for filtering and retrieval; they are not included in the embedding input.  This procedure is depicted in Fig.~\ref{fig:embeddings}.

To measure the local classification accuracy of papers in the embedding space, we now compute 
the $k$-Nearest-Neighborhood ($k$-NN) 
purity for the three types of embedding.  It is defined by
\begin{equation}
\text{Purity}@k = \frac{1}{N} \sum_{q=1}^{N} \frac{1}{k} \sum_{i=1}^{k} \mathbbm{1} \left[c\left(p_i^{(q)}\right) = c(q)\right]
\label{eq:knn_purity}
\end{equation}
Here, $N$ is the total number of query papers, $k$ is the neighborhood size, $p_i^{(q)}$ denotes the $i$-th nearest neighbor of query paper $q$ in the embedding space, and $c(\cdot)$ returns a paper's primary arXiv category. The indicator function $\mathbbm{1}[\cdot]$ equals 1 if the neighbor's category matches the query's category, and 0 otherwise. 

$\text{Purity}@k$, ranging from 0 to 1, thus measures the average of fraction of a paper's $k$ nearest papers in the embedding space that share its primary category.
We use cosine similarity as a metric to define the nearest. 

Considering $N$=$2500$, with papers drawn from hep-ph, hep-th, hep-ex, hep-lat, nucl-th, astro-ph, and gr-qc categories, we find that \textit{Qwen3}, \textit{Specter2} and \textit{Nomic} achieve nearly identical $k$-NN purity on 
our dataset. For the three embedding types: Purity$@10 \sim$0.66-0.67 and Purity$@5 \sim$0.68-0.69, see Tab.~\ref{tab:embedding-purity} for more details.
\begin{table}[H]  
  \centering
  \begin{tabular}{lcccc}
    \toprule
    Model & Parameters & Dim. & Purity@5 & Purity@10 \\
    \midrule
    \textit{Specter2} & 110 Million        & 768  & 0.68 & 0.66 \\
    \textit{Nomic}  & 137 Million     & 768  & 0.69 & 0.66 \\
     \textit{Qwen3} & 600 Million & 1024 & 0.69 & 0.67 \\
    \bottomrule
  \end{tabular} 
   \caption{Purity@$k$ (or $k$-NN purity) -- the fraction of the nearest-neighbors that share the query paper's primary category. The results are shown for three types of embedding models for sample sizes of $N$=$2500$. Here,  primary category means the first arXiv category.
 }
  \label{tab:embedding-purity}
\end{table}
Given that these quantitative gaps between models are small enough, it seems the purity alone does not justify preferring a larger general purpose embedding models in this case. Therefore, \textit{Specter2} seems suitable on practical and domain grounds. \textit{Specter2} being a lightweight model and that can be deployed efficiently on commodity hardware, makes it practical for local inference and large-scale indexing. 

Limiting only to primary categories hep-ph and astro-ph.HE, and averaging 
over the corpus ($N$=141,160 hep-ph, $N$=45,045 astro-ph primary category papers), we find
\begin{equation}
 \textit{\rm Purity@10:}\quad \textit{\rm  hep-ph=0.99}\,, \quad  \textit{\rm astro-ph.HE=0.97}. 
\end{equation}
Moreover, we find that $0.94$ of papers achieve Purity@10=1.  
This indicates 
\textit{Specter2} reliably locates the correct neighborhood.  

\begin{figure}[H]
    \centering
    \includegraphics[width=0.5\textwidth,height=0.3\textheight]{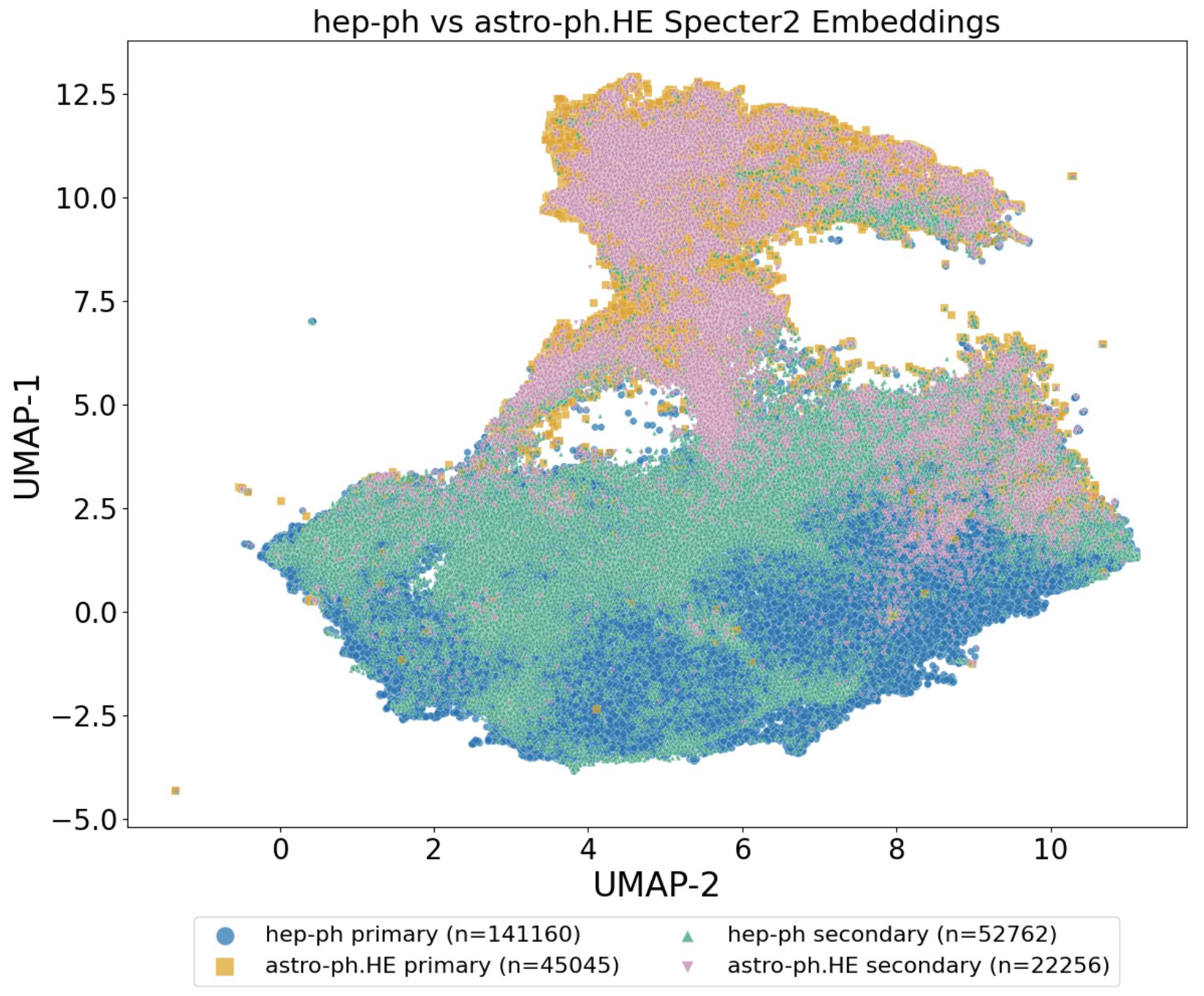}
    \caption{UMAP projection of \textit{Specter2} embeddings divided by primary and secondary arXiv categories. The hep-ph and astro-ph.HE as primary categories are occupying largely separate regions, shown in blue and orange, respectively. The hep-ph and astro-ph.HE as secondary categories are shown in green and magenta regions, respectively. The axes have no physical meaning. }
    \label{fig:umap}
\end{figure}
In Fig.~\ref{fig:umap}, we show the  Uniform Manifold Approximation and Projection (UMAP) of the full dataset containing papers with hep-ph and astro-ph.HE as primary or secondary categories. 
Two distinct clusters emerge, corresponding to the hep-ph (blue) and astro-ph.HE (red) primary populations, connected by a narrow bridge of overlapping points. Papers in which hep-ph or astro-ph.HE appear as a secondary category (green and yellow, respectively) are distributed across both primary clusters, with a visible concentration along the bridge, which is consistent with these cross-listed papers sharing thematic content with both fields.

To evaluate how well different embedding models separate subfields,
we constructed a category-balanced evaluation set drawn from our arXiv HEP 
corpus. We retained ten high-energy physics and astrophysics categories:
hep-ph, hep-th, hep-ex, hep-lat, nucl-th, gr-qc, astro-ph.HE,
astro-ph.SR, astro-ph.CO, and astro-ph.GA. Papers were sampled within each primary category as evenly as possible across publication years to minimize temporal bias. For the
comparative embedding analysis, we used 2500 papers per primary
category. Each paper was represented by its title + abstract text and
embedded using \textit{Specter2}, \textit{Nomic}, and \textit{Qwen3}. 
The resulting representations were then compared using UMAP
visualizations.
\begin{figure}[H] 
  \centering
    \includegraphics[width=1.0\textwidth]{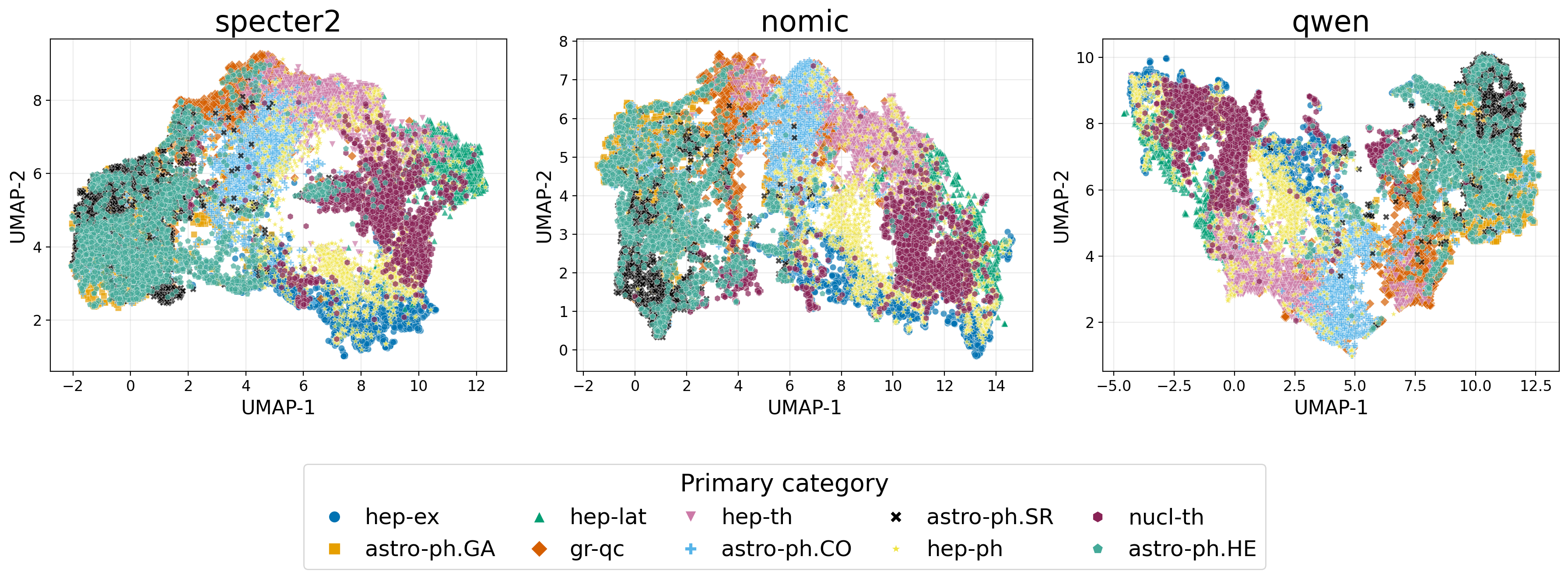}
  \caption{UMAP of papers from ten primary arXiv categories embedded with \textit{Specter2}, \textit{Nomic}, and \textit{Qwen3}. Points are colored by primary category, all models use the same stratified sample of size 2500. The axes have no physical meaning.}
 \label{fig:embeddings2}
\end{figure}
 This is shown in Fig.~\ref{fig:embeddings2},  
which  indicates clear differences in the extent 
to which the embedding models organize papers by primary category. 
\textit{Specter2} exhibits relatively well-defined category specific regions, 
although several neighboring categories show substantial overlap. 
This overlap is expected, as papers in the other categories are always 
cross-listed with either hep-ph or astro-ph.HE as a secondary category 
in our dataset, which is  also illustrated by the overlap between primary and secondary categories 
in Fig.~\ref{fig:dataset}.

\textit{Nomic} produces a similarly structured embedding space, with pronounced local 
clustering but noticeable mixing among several categories. \textit{Qwen3} also reveals 
a distinct category level structure, with some categories forming compact regions 
and others exhibiting greater overlap. Overall, the UMAP projections indicate
that all three models capture meaningful subfield structure, while the degree
and location of inter-category overlap vary across models.
From now on, we will always use \textit{Specter2} embedding for retrieval and 
synthesis.

\section{Hybrid Search Retrieval: Query $\rightarrow$ Top-\textit{k} papers}
\label{sec:retrieval}

In this section, we give details of our retrieval method, which is also 
shown in a flowchart in Fig.~\ref{fig:retrieval}.
The retrieval of semantically similar paper is refined through multistep 
procedure, as described in the following subsections. 

\usetikzlibrary{arrows.meta, positioning, fit, calc, shapes.geometric}

\begin{figure}[H]
\centering
\begin{tikzpicture}[
    font=\small,
    node distance=6mm and 10mm,
    box/.style={
        draw=blue!55!black, thick, rectangle, rounded corners=2pt,
        fill=blue!8, minimum width=2.6cm, minimum height=0.9cm,
        align=center, inner sep=3pt
    },
    input/.style={
        draw=blue!55!black, thick, cylinder, shape border rotate=90,
        fill=blue!8, aspect=0.25, minimum width=1.8cm, minimum height=1cm,
        align=center, inner sep=2pt
    },
    query/.style={
        draw=orange!70!black, thick, rectangle, rounded corners=2pt,
        fill=orange!12, minimum width=2.4cm, minimum height=0.9cm,
        align=center, font=\bfseries\small, inner sep=3pt
    },
    stagetitle/.style={font=\bfseries\small},
    arrow/.style={-{Stealth[length=2mm]}, thick, blue!55!black},
    stagebox/.style={draw, thick, inner sep=8pt, rounded corners=2pt}
]

\node[box] (embed)  at (2, 2.0) {Embed query\\(\textit{Specter2})};
\node[box, below=of embed] (hybrid) {Hybrid retrieve\\Dense + BM25 + RRF};
\node[box, below=of hybrid] (rerank) {Cross-encoder\\rerank};
\node[box, below=of rerank] (topk) {Top context\\papers};

\node[stagebox, fit=(embed)(hybrid)(rerank)(topk), label={[stagetitle]below:1) Retrieval}] (stage1) {};

\node[query] (query) at (2, 4.0) {User Query};

\node[input] (chroma) at (-2.6, 1.4) {Chroma\\vector DB};
\node[input] (bm25)   at (-2.6, 0.2) {BM25\\index};

\node[box, right=22mm of hybrid] (filter) {LLM-as-judge};
\node[box, below=of filter] (prompt) {Synthesis\\prompt};
\node[box, below=of prompt] (synth) {LLM synthesis\\+ citations};

\node[stagebox, fit=(filter)(prompt)(synth), label={[stagetitle]below:2) Synthesis}] (stage2) {};

\draw[arrow] (query) -- (stage1.north -| query) -- (embed.north);
\draw[arrow] (embed) -- (hybrid);
\draw[arrow] (chroma.east) -- (hybrid.west |- chroma.east) -- (hybrid.west);
\draw[arrow] (bm25.east) -- (hybrid.west |- bm25.east) -- (hybrid.west);
\draw[arrow] (hybrid) -- (rerank);
\draw[arrow] (rerank) -- (topk);

\draw[arrow] (topk.east) -- ++(0.8,0) |- (filter.west);

\draw[arrow] (filter) -- (prompt);
\draw[arrow] (prompt) -- (synth);

\end{tikzpicture}
\caption{\small Online retrieval pipeline used to obtain the candidate pool of papers for a 
given query. This is a multi-stage procedure consisting of dual retrieval: dense (via 
\textit{Specter2}) and sparse (via \textit{BM25}) -- fused via \textit{RRF}, followed by 
cross-encoder reranking. For synthesis, the LLM first judges the relevance of the retrieved 
papers and selects the final list, and then generates a concise report. }
\label{fig:retrieval}
\end{figure}
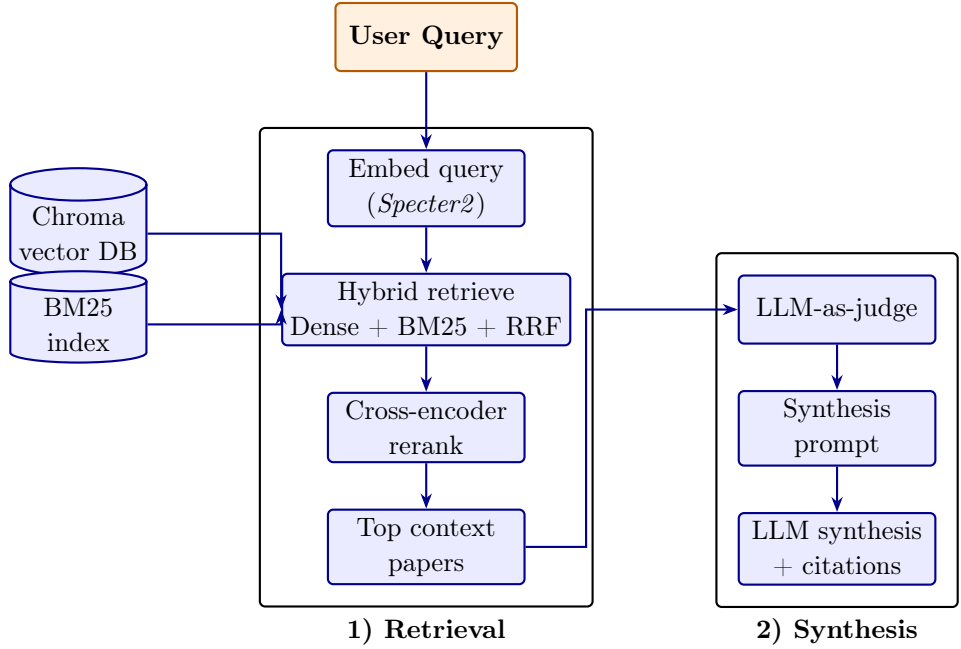

{\boldmath
\subsection{Dense retrieval}
}
\label{sec:query-to-topk}
The first step is to find top-$k$ similar  candidate papers 
for a given user query. 
For dense systematic retrieval, the query needs to be also embedded, for which we use two complementary models.  
First, we use  \textit{Specter2} model along with 
\textit{adhoc query} adapter \cite{specter2} to generate an embedding of query. 
Subsequently, we compare this query embedding against precomputed embeddings, using \textit{Specter2} along with \textit{proximity} adapter, of paper 
stored in offline vector database. After initial retrieval, candidate papers are ranked using cosine 
similarity defined by
\begin{equation}
cos(q, d) =
\frac{q \cdot d}{\|q\| \, \|d\|}.
\end{equation}
Here, $d$ and $q$ denote the document and query vectors, and ${\|v\|}$ stands for the Euclidean norm of a vector $v$.

{\boldmath
\subsection{Sparse retrieval}
}
{For sparse lexical retrieval, we use \emph{Best Matching 25} (BM25) model \cite{robertson2009bm25}}. It ranks documents on the basis of three factors: how often query terms appear in each paper's abstract and title, how rare those terms are across the whole HEP-Astro corpus (rare terms count more than common ones), and the document's length. 
In BM25, the latter is required  
because otherwise long documents can be artificially favoured due to higher 
probability of occurrence of term.
BM25 is an effective method for keyword matching.

By combining dense and sparse paradigms, our retrieval leverages 
both semantic similarity and lexical relevance to improve the quality of candidate papers.

\subsection{Reciprocal rank fusion}
In order to combine the outputs from dense and sparse retrievals, we use \emph{Reciprocal Rank Fusion} (RRF) \cite{cormack2009rrf},  since scores 
produced by these two models are not directly comparable. The RRF  
aggregates ranks instead of raw scores differing in scales. 
For each candidate paper $d$, the fused score is defined by
\begin{equation}
{\rm RRF}(d) = 
\frac{1}{k+{\rm rank}_{\rm dense}(d)} + \frac{1}{k+{\rm rank}_{\rm sparse}(d)}.
\end{equation}
Here, ${\rm rank}_r(d)$ denotes the rank of paper $d$ in the list returned by retriever $r$. 
We use $k=60$ which is a standard RRF configuration \cite{cormack2009rrf}.  
The candidate papers are then sorted according to their fused scores, producing a unified ranking.

{\boldmath
\subsection{Cross-encoder reranker}
}
The fused candidate list is subsequently refined using \textit{ms-marco-MiniLM-L-6-v2}, a cross-encoder reranker trained on the MS MARCO passage-ranking task \cite{ms-marco-MiniLM-L-6-v2,nogueira2019passage}. 

Given a query $q$ and a candidate paper $d$, the model jointly encodes the query and the document to predict a relevance score
\begin{equation} 
r(q,d) = f_{\theta}\!\left([{\tt CLS}] \; q \; [{\tt SEP}] \; d \; [{\tt SEP}]\right), 
\end{equation}
where $f_{\theta}$ represents a transformer-based model having parameters $\theta$.
{\tt [CLS]} is classification token prepended to input sequence, and {\tt [SEP]} is separation token, typically needed in a transformer model.   
Unlike bi-encoders, which independently encode queries and documents before comparing their embeddings, cross-encoders jointly attend to both inputs, enabling fine-grained interactions between query and document tokens. As a result, they provide more accurate relevance estimates and are well suited to rerank a relatively small set of retrieved candidates. The complete retrieval pipeline is illustrated in Fig.~\ref{fig:retrieval}.

\section{Benchmark and Evaluation}
\label{sec:bench-eval}
To evaluate retrieval quality, we constructed two benchmark datasets, one for hep-ph and one for astro-ph.HE. The two differ slightly in nature, which adds variety to the evaluation, as described below.
These will be made publicly available on Hugging Face.

\subsection{Benchmarks for hep-ph and astro-ph.HE}
\begin{figure}[H] 
  \centering
    \includegraphics[width=1.0\textwidth]{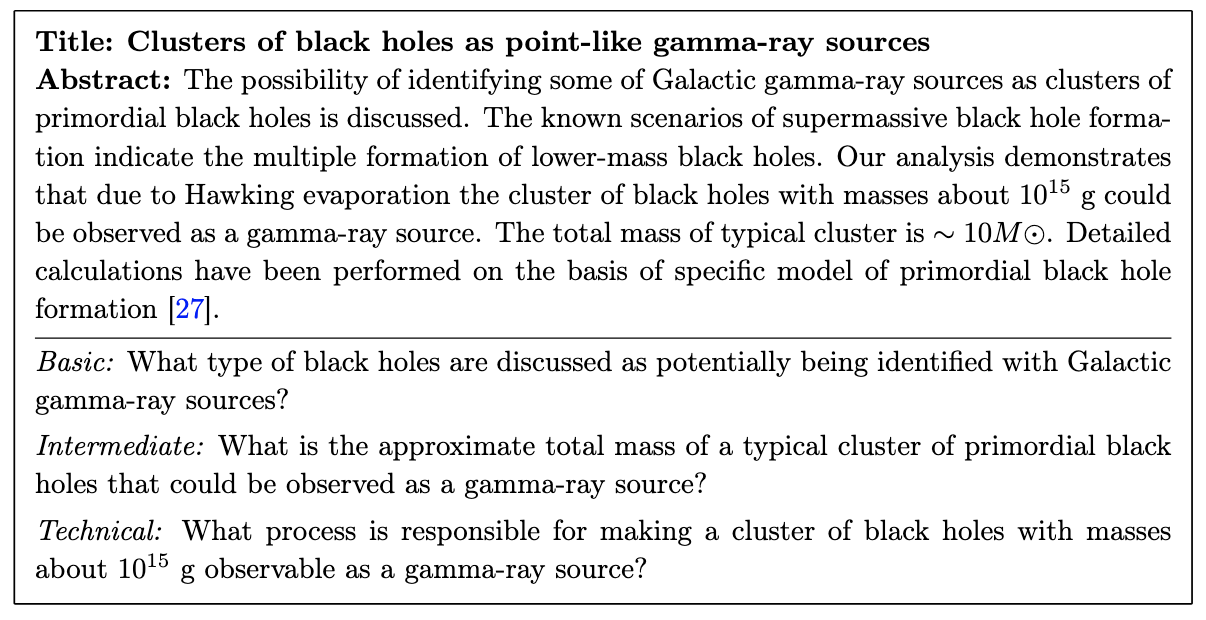}
    \caption{\small An example from the astro-ph.HE evaluation benchmark (source paper \cite{Belotsky_2011}), consisting of an abstract and 
three question types (basic, intermediate, technical) used to evaluate 
different retrieval and reasoning capabilities.}
\label{fig:benchmark-astro-ph}
\end{figure}

For each hep-ph paper's abstract in the evaluation dataset, we generated three types of synthetic questions, each probing a different aspect of the abstract. These questions were generated using the \textit{Llama4-Maverick-17B-Instruct} model \cite{meta2025llama4} on Amazon Bedrock. For hep-ph, three question types are defined as follows:
\begin{itemize}
    \item \textbf{Main findings}: a question about the results reported in the abstract.
    \item \textbf{Claim verification}: a question that probes a specific claim made in the abstract.
    \item \textbf{Vague exploratory}: a broader and  more abstract question related to the paper's general topic.
\end{itemize}
The astro-ph.HE benchmark is slightly different. Three questions type for astro-ph.HE is as follows:
\begin{itemize} \item \textbf{Basic:} an object, instrument, or source named in the opening of the abstract. \item \textbf{Intermediate:} one measured feature, count, or implication from the body of the abstract. \item \textbf{Technical:} the quantitative conclusion, or the model requirement that explains it. 
\end{itemize}
\begin{figure}[t] 
  \centering
    \includegraphics[width=1.0\textwidth]{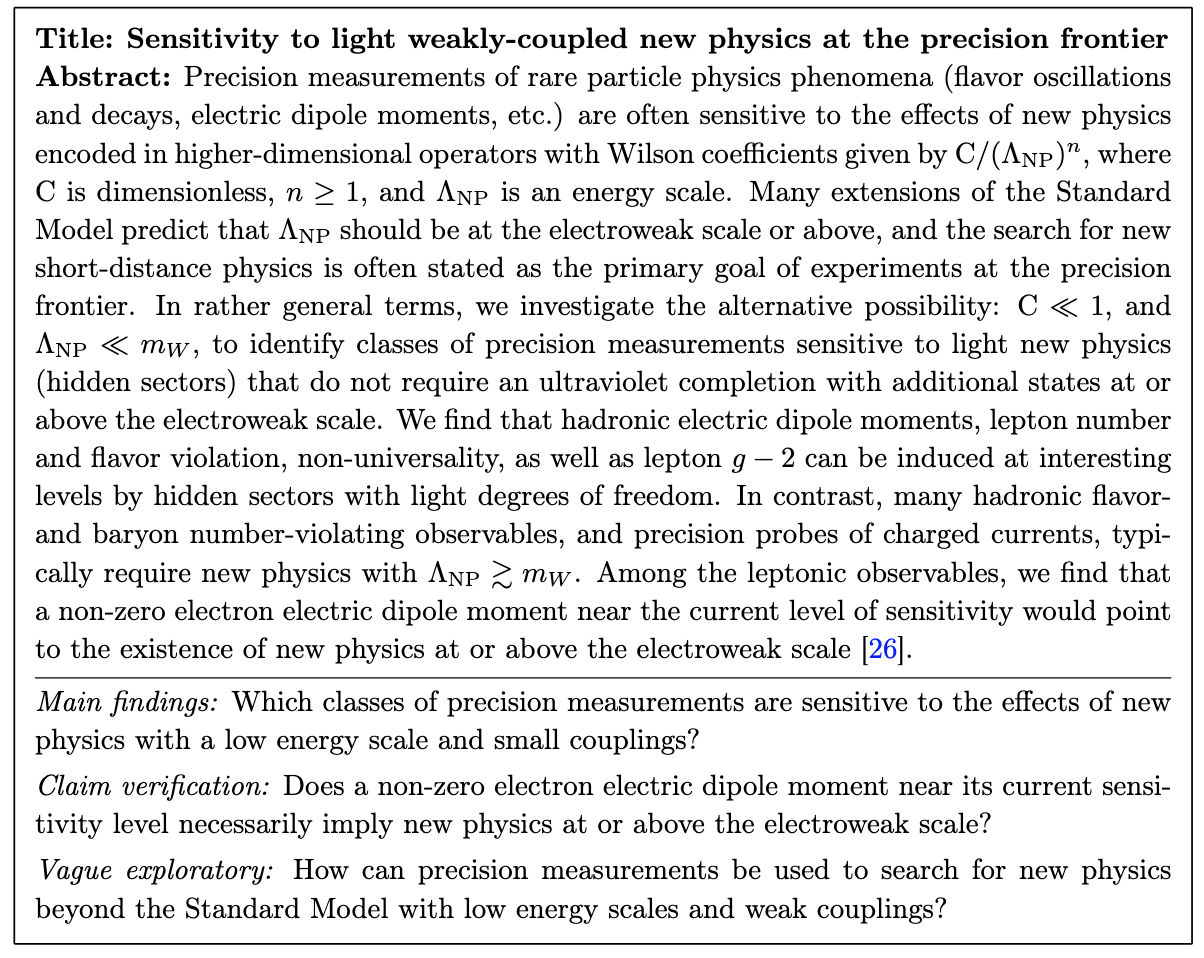}
  \caption{\small An example from the hep-ph evaluation benchmark (source paper \cite{LeDall:2015ptt}), consisting of an abstract and 
three question types (main findings, claim verification, vague exploratory) used to evaluate 
different retrieval and reasoning capabilities.}
\label{fig:benchmark-hep-ph}
\end{figure}
Examples of each question type, along with the corresponding  abstract, are shown for hep-ph and astro-ph.HE in Fig.~\ref{fig:benchmark-hep-ph} and  \ref{fig:benchmark-astro-ph}, respectively. 
The benchmark datasets consist of 2040 samples for hep-ph and 2376 samples for astro-ph.HE. 
They were created by prompting the \textit{Llama4-Maverick-17B-Instruct} model with the title and 
abstract of each paper, together with three example questions of each type for the respective 
category (hep-ph or astro-ph.HE). 

We then evaluate whether the pipeline can correctly retrieve the source paper given a 
query from this benchmark.

\subsection{Evaluation}
In this section, we discuss the evaluation of our pipeline based on the 
two benchmarks detailed in the previous subsection. For this purpose, we employ binary $\text{Recall@}k$ 
and Mean Reciprocal Rank (MRR) metrics. The $\text{Recall@}k$ measures, whether the 
relevant single paper appears anywhere within top-$k$ results, and the MRR measures ranking 
quality which is not captured by recall.

For query $q$ from a dataset of total $Q$ queries in the benchmark, the 
$\text{Recall@}k$ is defined as
\begin{equation}
\text{Recall@}k = \frac{1}{|Q|}\sum_{q \in Q} \frac{|R_q \cap T_k(q)|}{|R_q|}.
\end{equation}
Here, $R_q$ is the set of true papers for query $q$, but, since in our benchmark  
we associate only a single paper for each query, this implies $|R_q|=1$.   $T_k(q)$ denotes 
the set of top-$k$ retrieved papers 
and $|Q|=2404$ for hep-ph and $|Q|=2376$ for astro-ph.HE. 

Suppose that for query $q$, the rank of correctly retrieved paper comes out to be
${\text{rank}_q}$, the MRR is then given by
\begin{equation}
\text{MRR} = \frac{1}{|Q|}\sum_{q \in Q} \frac{1}{\text{rank}_q}.
\end{equation}
In Tab.~\ref{tab:retrieval-results-hep-ph} and \ref{tab:retrieval-results-astro}, 
we present the evaluation results on recall and MRR for the 
hep-ph and astro-ph.HE papers for $k=3,5$ and $10$. The results are divided by the 
 type of queries and the method (dense only, hybrid, hybrid+rerank 
 as discussed in Sec.~\ref{sec:query-to-topk}) used in the retrieval process. 
\begin{table}[H]
\centering
\small
\begin{tabular}{ll ccc ccc}
\toprule
\multirow{2}{*}{\textbf{Method}} & \multirow{2}{*}{\textbf{Category}} & \multicolumn{3}{c}{\textbf{Recall$@k$}} & \multicolumn{3}{c}{\textbf{MRR$@k$}} \\
\cmidrule(lr){3-5} \cmidrule(lr){6-8}
 & & @3 & @5 & @10 & @3 & @5 & @10 \\
\midrule
\multirow{3}{*}{Dense}
 & Claim Verification & 0.305 & 0.343 & 0.391 & 0.263 & 0.272 & 0.279 \\
 & Main Finding       & 0.291 & 0.330 & 0.381 & 0.248 & 0.257 & 0.264 \\
 & Vague Exploratory  & 0.055 & 0.071 & 0.100 & 0.042 & 0.045 & 0.049 \\
\midrule
\multirow{3}{*}{Hybrid}
 & Claim Verification & 0.629 & 0.757 & 0.893 & 0.498 & 0.527 & 0.546 \\
 & Main Finding       & 0.594 & 0.698 & 0.836 & 0.474 & 0.498 & 0.516 \\
 & Vague Exploratory  & 0.149 & 0.204 & 0.300 & 0.113 & 0.125 & 0.138 \\
\midrule
\multirow{3}{*}{Hybrid + Rerank}
 & Claim Verification & 0.917 & 0.937 & 0.958 & 0.856 & 0.861 & 0.864 \\
 & Main Finding       & 0.865 & 0.896 & 0.923 & 0.800 & 0.807 & 0.811 \\
 & Vague Exploratory  & 0.332 & 0.374 & 0.422 & 0.272 & 0.282 & 0.288 \\
\bottomrule
\end{tabular}

\vspace{4pt}
\footnotesize
\textit{Note:} $n = 2404$ for each category.
\caption{Retrieval performance by method and question category for hep-ph papers.}
\label{tab:retrieval-results-hep-ph}
\end{table}
Across both the hep-ph (Tab.~\ref{tab:retrieval-results-hep-ph}) and astro-ph.HE 
(Tab.~\ref{tab:retrieval-results-astro}) benchmarks, retrieval performance improves 
consistently and substantially as the different components are added to the pipeline: Dense 
retrieval alone performs weakest, Hybrid (sparse + dense fusion) improves markedly 
over Dense, and Hybrid~+~Rerank achieves the best results across every metric and 
category. 

For hep-ph, Recall@10 on Claim Verification rises from $0.391$ (Dense) to 
$0.893$ (Hybrid) to $0.958$ (Hybrid~+~Rerank), and a similar trend holds for MRR$@ k$, 
indicating that reranking not only retrieves the correct paper more often but also 
ranks it closer to the top. 

Vague Exploratory queries are consistently the hardest 
category across both methods and both domains, with Recall$@10$ of only $0.100$ 
(Dense) and $0.300$ (Hybrid) for hep-ph, and even Hybrid~+~Rerank reaching just 
$0.422$ -- far below the Claim Verification and Main Finding scores, reflecting 
the inherent difficulty of matching broad, under specified queries to a single 
correct source paper. 

\begin{table}[H]
\centering

\small
\begin{tabular}{ll ccc ccc}
\toprule
\multirow{2}{*}{\textbf{Method}} & \multirow{2}{*}{\textbf{Difficulty}} & \multicolumn{3}{c}{\textbf{Recall$@k$}} & \multicolumn{3}{c}{\textbf{MRR$@k$}} \\
\cmidrule(lr){3-5} \cmidrule(lr){6-8}
 & & @3 & @5 & @10 & @3 & @5 & @10 \\
\midrule
\multirow{3}{*}{Dense}
 & Basic        & 0.199 & 0.234 & 0.294 & 0.159 & 0.167 & 0.176 \\
 & Intermediate & 0.328 & 0.379 & 0.449 & 0.268 & 0.280 & 0.289 \\
 & Technical    & 0.357 & 0.398 & 0.463 & 0.300 & 0.310 & 0.318 \\
\midrule
\multirow{3}{*}{Hybrid}
 & Basic        & 0.458 & 0.562 & 0.712 & 0.357 & 0.381 & 0.401 \\
 & Intermediate & 0.658 & 0.757 & 0.872 & 0.534 & 0.556 & 0.572 \\
 & Technical    & 0.656 & 0.761 & 0.885 & 0.541 & 0.565 & 0.582 \\
\midrule
\multirow{3}{*}{Hybrid + Rerank}
 & Basic        & 0.787 & 0.832 & 0.869 & 0.724 & 0.734 & 0.739 \\
 & Intermediate & 0.891 & 0.921 & 0.947 & 0.830 & 0.837 & 0.840 \\
 & Technical    & 0.885 & 0.913 & 0.941 & 0.821 & 0.828 & 0.831 \\
\bottomrule
\end{tabular}

\vspace{4pt}
\footnotesize
\textit{Note:} $n = 2376$ for each difficulty level.
\caption{Retrieval performance by method and difficulty level for astro-ph.HE papers.}
\label{tab:retrieval-results-astro}
\end{table}

For astro-ph.HE, a similar pattern holds across difficulty 
levels (Basic, Intermediate, Technical), with Hybrid~+~Rerank achieving Recall$@10$ 
above $0.86$ for all levels, though interestingly Basic queries perform slightly 
worse than Intermediate and Technical ones under Dense and Hybrid retrieval, 
suggesting that overly simple queries may lack the distinctive keywords needed for 
effective lexical/dense matching -- an effect that reranking largely corrects.

\section{Synthesizing HEP Review Reports with RAG}
\label{sec:generation}
We finally demonstrate the application of our retrieval pipeline for the downstream task of generating short review reports with appropriate references for a given query. However, although our benchmarks,  each containing a single target paper, show that our pipeline achieves high recall and MRR, report generation requires feeding multiple relevant references to an LLM. Moreover, the top-$k$ retrieved papers often include articles that are semantically similar to the query but still topically off-target. 

To address this, we first retrieve a large candidate pool of papers ($K=100$) for each query, and then use an LLM -- as a judge, to select a final refined list ($F$) of references based on relevance.  For the purpose of relevance check as well as generation we use \textit{Gemma-4-E4B-IT} model with 4.5B effective parameters from Google DeepMind. Throughout, we will refer this as \textit{Gemma4} model. 

Further details are given in the following subsection.

\subsection{LLM-as-a-Judge for relevance}
\label{sec:llm-judge}
The relevance filtering step in our pipeline follows the ``LLM-as-a-judge'' method \cite{zheng2023judging},  in which an LLM is used in place of a human to 
assess relevance between initially retrieved candidates using hybrid search and cross-encoder reranking.
In our setting, the LLM judge is given the query together with a numbered
list of candidate papers (title and abstract) and is
prompted to select the subset of $F$ most relevant papers to the query.

As discussed before, this design choice is deliberate: rather than relying solely on
embedding similarity or lexical overlap rankings, we use the LLM's judgment
as an additional relevance filter, exploiting its capacity for contextual
and semantic reasoning about relevance that rankers lack.
For relevance filtering, the prompt given to LLM is:
\begin{figure}[H]
\centering
\begin{tcolorbox}[
    colback=white,
    colframe=black,
    boxrule=0.6pt,
    arc=0pt,
    width=0.95\textwidth,
    left=4pt, right=4pt, top=4pt, bottom=4pt
]
\small
\textbf{Relevance Prompt:}
You are a research assistant. From the numbered list of papers below, 
select the $F$ most relevant to the query. Return ONLY the numbers as a 
comma-separated list, e.g.: 1,3,5,7. {\tt Query:q, candidate papers, Selected numbers:}
\end{tcolorbox}
\caption{Prompt given to \textit{Gemma4} model to select most relevant papers.}
\label{fig:prompt}
\end{figure}
 
We note that when acting as judge, LLMs are known to exhibit certain biases, including position, verbosity, and self-enhancement bias~\citep{zheng2023judging}. Self-enhancement bias, where an LLM favours outputs generated by models from its own family, is not directly applicable to our setting, since the \textit{Gemma4} evaluates retrieved original arXiv content rather than LLM generated text.

Regarding positional bias, we empirically tested the sensitivity of relevance selections 
to candidate ordering of $K$ retrieved papers by running the filtering prompt across 
different orderings of the same candidate set: the original order, its reverse, and five randomly reshuffled 
permutations (fixed random seed for reproducibility). We found selections to be highly sensitive to 
ordering, with a mean pairwise Jaccard \footnote{The Jaccard Index between two sets $A$ and $B$ is defined as $J(A,B) = \frac{|A \cap B|}{|A \cup B|}$.} varying a lot across the $\binom{7}{2}=21$ pairwise comparisons of permuted runs. We therefore adopt the union of papers selected by LLM judge across these seven permuted runs of retrieved papers as our final candidate set for synthesis, which improves robustness to positional bias at the cost of a larger and query-dependent reference set size .

Concerning verbosity bias, we computed the token length of title and abstract 
using \textit{Gemma4} tokenizer. 
We find that it is moderately consistent across arXiv articles in our dataset, with mean = 241, median = 223, and standard deviation = 111, suggesting most inputs given to \textit{Gemma4} model for synthesis fall within a comparable range. Therefore, we expect any verbosity bias effects from input length alone to be limited, though we cannot rule out bias arising from other differences across abstracts and titles.

\subsection{Synthesizing reports with LLM}

Starting from a query on a given topic in HEP-Astro and ending in a concise report, we combine 
all the steps: retrieval, relevance check, and report synthesis using an LLM 
(\textit{Gemma4}) into Algorithm~\ref{alg:recursive-rag}. See also Fig.~\ref{fig:retrieval}.
 Using this algorithm, we are able to readily generate a report on any topic within either hep-ph or astro-ph.HE. In the current implementation, the query is a direct input to retrieval system and must therefore be well-defined. Furthermore, the synthesis prompt provided to the \textit{Gemma4} model is:
\begin{figure}[H]
\centering
\begin{tcolorbox}[
    colback=white,
    colframe=black,
    boxrule=0.6pt,
    arc=0pt,
    width=0.95\textwidth,
    left=4pt, right=4pt, top=4pt, bottom=4pt
]
\small
\textbf{Synthesis Prompt:}
You are a research assistant. Write a concise synthesis report (4-5 paragraphs) 
summarizing the key themes, findings, and connections across the papers below, 
in the context of the query only. Properly cite papers at relevant places in the report. 
{\tt Query:q, synthesis context, Synthesis:}
\end{tcolorbox}
\caption{Prompt given to \textit{Gemma4} model for report synthesis.}
\label{fig:promptsynthesis}
\end{figure}
It is used in step 3 of the Algorithm~\ref{alg:recursive-rag}.
For illustration purposes, we generate only a short report of 4-5 paragraphs in length, however, more comprehensive reports can also be generated by increasing the number of final papers ($F$) selected by the LLM judge and initial papers $(K)$ from the hybrid retrieval.
\begin{algorithm}[H]
\caption{Retrieval-Augmented Report Synthesis}
\label{alg:recursive-rag}
\begin{algorithmic}[H]
\Require A user query $q$, initial candidate size $K$, final selection size $F$, number of random shuffles $N$
\Ensure A concise report synthesis with citations.

\Statex \textbf{Step 1: Initial Retrieval}
\State --- Retrieve the top $100$ candidate papers ($K$) for a query using a hybrid search that combines dense embedding similarity and BM25 matching, fuse the two rankings, and reranks the merged list with a cross-encoder model.

\Statex \textbf{Step 2: LLM-Based Relevance Filtering with Position Shuffle}
\State --- Construct a set of $N+2$ orderings of the $K$ candidates: the original order after cross-encoder, its reverse, and $N$=$5$ independent random permutations. For each ordering $o$, present the $K$ candidates to the \textit{Gemma4} model as a numbered list following ordering $o$, each entry showing the paper's title and abstract.
    \State --- Using a prompt, given in Fig.~\ref{fig:prompt}, ask the \textit{Gemma4} model to select the $F$ papers most relevant to the query and to return their list numbers.
\State --- Form the filtered paper set as the union $\bigcup_o S_o$ of the selected sets across all orderings.

\Statex \textbf{Step 3: Synthesis Generation with LLM}
\State --- Use the synthesis prompt, given in Fig.~\ref{fig:promptsynthesis}, containing the query together with the title and abstract of each paper in the union.
\State --- Ask the \textit{Gemma4} model to write a short report that addresses the query with citations.

\State --- Return the synthesis report and the union paper set $\bigcup_o S_o$.
\end{algorithmic}
\end{algorithm}

\subsection{Example Reports}
We present two example reports synthesized by our algorithm, one each for hep-ph and astro-ph.HE. For brevity, we include only one example report per field here; additional examples will be made available on the GitHub page of the repository.

To generate a short report on hep-ph, we selected a topic aligned with the area of expertise of one author, enabling expert evaluation of the resulting report. Starting from an initial retrieval of $K$=$100$ candidate papers, the LLM-judge relevance check selected $F$=$28$ papers through the union of reshuffled runs. These papers were then passed as context to the \textit{Gemma4} model, together with a synthesis prompt, to generate the final report. The resulting report is shown in Fig.~\ref{fig:report-smeft}.
\begin{figure}[H]
\centering
\begin{tcolorbox}[
    colback=white,
    colframe=black,
    boxrule=0.6pt,
    arc=0pt,
    width=0.95\textwidth,
    left=4pt, right=4pt, top=4pt, bottom=4pt
]
\small

\textbf{Query:}
How to calculate anomalous dimensions in SMEFT to compute renormalization group running effects?

\vspace{1.5mm}
\hrule
\vspace{1.5mm}

\textbf{Synthesis Report:}
The calculation of anomalous dimensions in the Standard Model Effective Field Theory (SMEFT) is crucial for consistently computing renormalization group (RG) running effects, which ensures that predictions for physical observables are independent of the arbitrary renormalization scale ($\mu$) [arXiv:2601.19974]. These anomalous dimensions govern how the Wilson coefficients of higher-dimension operators evolve between different energy scales. The calculation methodologies employed are diverse, ranging from traditional loop calculations to more efficient, amplitude-based approaches. Specifically, on-shell methods have proven to be an exceptionally efficient technique for deriving one-loop anomalous dimensions of higher-dimension operators directly from tree-level amplitudes, bypassing explicit loop calculations [arXiv:2005.07129, arXiv:2005.10261].

The complexity of these calculations scales with the dimension of the operators and the loop order. For dimension-six operators, one-loop anomalous dimension matrices have been calculated, detailing the contributions from various Standard Model couplings, such as the Higgs self-coupling ($\lambda$) and Yukawa couplings (y) [arXiv:1308.2627]. Furthermore, the development of higher-order calculations is ongoing; for instance, two-loop anomalous dimensions have been described using on-shell methods [arXiv:2005.12917], and the full two-loop RG equations for specific sectors, like baryon-number-violating operators, have been computed [arXiv:2510.08682].

The application of these calculated anomalous dimensions is central to phenomenology. RG running allows for the consistent combination of collider observables measured at different energy scales, such as those in global fits [arXiv:2212.05067]. This running is necessary to account for operator mixing, which is a key feature of SMEFT, ensuring scale independence in Next-to-Leading Order (NLO) calculations [arXiv:2601.19974]. Specific applications include studying the impact of RG running on Higgs production processes at the LHC [arXiv:2406.06670, arXiv:2409.19578] and analyzing flavor-violating effects induced by Yukawa running [arXiv:2005.12283].

Beyond the standard SMEFT, the framework extends to theories incorporating new physics, such as sterile neutrinos ($\nu$SMEFT) [arXiv:2411.07220] or axion-like particles (ALPs) [arXiv:2105.01078]. In these extended scenarios, the anomalous dimensions must account for the new particle interactions, including Yukawa couplings [arXiv:2407.16751]. Moreover, the framework is being extended to higher dimensions, with theoretical work addressing dimension-seven [arXiv:2310.11055] and dimension-eight operators [arXiv:2203.06771, arXiv:2409.10305], necessitating the computation of corresponding RG evolution matrices [arXiv:2408.15378].
\end{tcolorbox}
\caption{\small Synthesis report on hep-ph generated by our RAG pipeline from $F$=$28$ papers selected out of $K$=$100$ retrieved candidates.}
\label{fig:report-smeft}
\end{figure}

\begin{figure}[H]
\centering
\begin{tcolorbox}[
    colback=white,
    colframe=black,
    boxrule=0.6pt,
    arc=0pt,
    width=0.95\textwidth,
    left=4pt, right=4pt, top=4pt, bottom=4pt
]
\small

\textbf{Query:}
How does the VERITAS constrain WIMP dark-matter annihilation cross sections using gamma-ray observations?

\vspace{1.5mm}
\hrule
\vspace{1.5mm}

\textbf{Synthesis Report:}
The search for Weakly Interacting Massive Particle (WIMP) dark matter (DM) annihilation is a primary goal in astrophysics, leveraging the prediction that DM self-annihilation in regions of high DM density can produce detectable Very High Energy (VHE) gamma rays [arXiv:1110.4358, arXiv:1110.6615]. Dwarf spheroidal galaxies (dSphs) are considered prime targets for these indirect searches because they are expected to possess high DM content while exhibiting low astrophysical backgrounds [arXiv:1307.8367, arXiv:1210.4961]. VERITAS, as an imaging atmospheric Cherenkov telescope array, is specifically designed to detect these VHE gamma rays, which can have energies up to the DM particle mass [arXiv:2407.16518].

VERITAS constrains WIMP annihilation cross sections by setting upper limits on the observed gamma-ray flux from these dSphs. The absence of a statistically significant gamma-ray signal in observations of specific dSphs, such as Segue 1 [arXiv:1110.4358], Draco, Ursa Minor, and Willman 1 [arXiv:0902.4750], allows researchers to derive conservative bounds on the DM annihilation cross-section [arXiv:0902.4750]. Furthermore, joint statistical analyses combining data from multiple dSphs have been employed to enhance sensitivity and tighten these constraints on the annihilation cross section [arXiv:1703.04937, arXiv:2407.16518].

The methodology involves comparing the expected gamma-ray flux, derived from theoretical models assuming WIMP annihilation, against the upper limits established by VERITAS observations [arXiv:1210.4961]. These constraints are crucial for testing various particle physics models, particularly those predicting WIMPs in the mass range of 50 GeV to over 10 TeV [arXiv:1307.8367, arXiv:0910.4563]. While VERITAS has focused on dSphs, related studies also examine other targets, such as the Galactic Center [arXiv:2309.12403], to constrain DM properties, although dSphs remain a key focus for robust cross-section limits [arXiv:2003.13482].

In summary, VERITAS constrains WIMP annihilation cross sections by performing null searches for VHE gamma rays originating from dSphs. By utilizing the high DM density of these galaxies and the sensitivity of the IACT array, researchers translate non-detections into stringent upper limits on the annihilation cross section, thereby ruling out certain parameter spaces for WIMP models [arXiv:1703.04937, arXiv:1210.4961].
\vspace{1mm}

\end{tcolorbox}
\caption{\small Synthesis report on astro-ph.HE generated by our RAG pipeline from $F$=$25$ papers selected out of $K$=$100$ retrieved candidates.}
\label{fig:reprot-astro}
\end{figure}

We have checked that the above hep-ph report given in Fig.~\ref{fig:report-smeft}, which is based on the titles and abstracts of 17 research papers, is coherent, correct, and free of hallucinations. Furthermore, we verified that each statement in the report is accurate and is supported by the corresponding cited paper. The report covers seminal work on the topic, while also capturing recent progress, such as the use of on-shell methods for computing anomalous dimensions as well as phenomenological impact of renormalization group running effects. We therefore conclude that the report and the papers it cites provide a very good starting point for work on the computation of anomalous dimensions in the SMEFT and beyond. We analyses several other such report and found a 
similar level of quality.  

For astro-ph.HE, we selected a topic aligned with the area of expertise of another author, again enabling expert evaluation of the resulting report. From an initial retrieval of $K$=$100$ candidate papers, the union of the selections made by the LLM judge over several reshuffled runs yielded $F$=$25$ papers. These papers were then passed to \textit{Gemma4} as context, together with a synthesis prompt, to generate the synthesis report shown in Fig.~\ref{fig:reprot-astro}.

Similarly, the synthesis report generated for the astro-ph.HE query, as shown in Fig.~\ref{fig:reprot-astro},  provides a clear and coherent overview of how VERITAS searches for WIMP dark matter through the indirect detection of very-high-energy (VHE) gamma rays. Interestingly, the report connects the theoretical motivation for the annihilation of WIMP with the observational strategy. For example, dwarf spheroidal galaxies (dSphs) were selected as targets because of their high DM mass content and low astrophysical background. We explicitly checked that the cited references and the factual claims of no dark matter detection are accurate. 
More such reports and their expert analyses will be provided on the GitHub repository. 

\section{Conclusion and Future Directions}
\label{sec:conclude}

AI has already affected the way we conduct research in HEP and Astroparticle physics. A researcher's day-to-day activities now include code generation, finding references, and bouncing ideas off LLMs such as Claude, ChatGPT, and Gemini. These general-purpose LLMs, though extremely powerful, also have limitations such as knowledge cutoffs, subscription requirements, and safety concerns. For these reasons, it becomes extremely important to develop a small, open-source AI system that can serve a specific domain such as high-energy physics without suffering from the above problems.

RAG is a powerful AI framework that allows LLMs to generate text well-grounded in an external knowledge source.
In this work, we present an open-source and modular RAG pipeline for finding and synthesizing HEP and Astro  research papers, focusing on the hep-ph and astro-ph.HE categories.
The focus on specialized subdomains hep-ph and astro-ph.HE addresses an identified gap in existing RAG literature. 

We embedded about 230K arXiv articles using the \textit{Specter2} embedding model and stored them in a Chroma vector database. We then retrieve semantically similar papers for a given query using a hybrid-search RAG approach and rerank them using the \textit{ms-marco-MiniLM-L-6-v2} cross-encoder model. We found that this RAG pipeline achieves very high recall and MRR, evaluated using two synthetic benchmarks in hep-ph and astro-ph.HE.

Post retrieval, a large number of papers are further refined based on their relevance to the query with the help of the Google DeepMind's \textit{Gemma4} model, which is open-source, used as an LLM-as-a-judge ensuring bias mitigation. The final set of papers is then used as context for the \textit{Gemma4} model to generate a short (4–5 paragraphs) review with citations. 
Based on the author's domain knowledge of hep-ph and astro-ph.HE, respectively, it is found that the resulting reports to be coherent, technically sound, and well grounded in the final list of papers.

As an application, such a method can be used by researchers, reviewers, or editors to quickly get up to speed on the current literature, conduct a gap and novelty analysis, or find new directions to work on. We see many possibilities to improve our method, such as providing the full text of papers as context to LLM for synthesis, query decomposition prior to retrieval and incorporating citation information of source papers into retrieval and synthesis.  Finally, developing a rigorous and scalable method for evaluating the domain-specific reports synthesized by our pipeline, without requiring human involvement, would be extremely valuable.

\acknowledgments

This work was financially supported by the Natural
Sciences and Engineering Research Council of Canada (J.K.). 
This research was enabled in part by support provided the Digital Research Alliance of Canada
(www.alliancecan.ca). Computations were performed
on the Trillium supercomputer at the SciNet \cite{Loken_2010} HPC
Consortium. SciNet is funded by Innovation, Science and Economic Development Canada; the Digital Research Alliance of Canada; the Ontario Research Fund:
Research Excellence; and the University of Toronto.

\bibliography{sample}

\providecommand{\href}[2]{#2}\begingroup\raggedright\begin{thebibliography}{10}

\bibitem{Richmond:2025lzg}
P.~Richmond, C.~Papageorgakis, V.~Niarchos, B.~Chowdhury and P.~Agarwal,
  \emph{{FeynTune: large language models for high-energy theory}},
  \href{https://doi.org/10.1088/2632-2153/ae47bb}{\emph{Mach. Learn. Sci.
  Tech.} {\bfseries 7} (2026) 025012}
  [\href{https://arxiv.org/abs/2508.03716}{{\ttfamily 2508.03716}}].

\bibitem{nguyen-etal-2023-astrollama}
T.D.~Nguyen, Y.-S.~Ting, I.~Ciuca, C.~O{'}Neill, Z.-C.~Sun, M.~Jab{\l}o{\'n}ska
  et~al., \emph{{A}stro{LL}a{MA}: Towards specialized foundation models in
  astronomy},  in \emph{Proceedings of the Second Workshop on Information
  Extraction from Scientific Publications}, T.~Ghosal, F.~Grezes, T.~Allen,
  K.~Lockhart, A.~Accomazzi and S.~Blanco-Cuaresma, eds., (Bali, Indonesia),
  pp.~49--55, Association for Computational Linguistics, Nov., 2023,
  \href{https://doi.org/10.18653/v1/2023.wiesp-1.7}{DOI}.

\bibitem{rag}
P.~Lewis, E.~Perez, A.~Piktus, F.~Petroni, V.~Karpukhin, N.~Goyal et~al.,
  \emph{Retrieval-augmented generation for knowledge-intensive nlp tasks},
  2021.

\bibitem{gao2024retrievalaugmentedgenerationlargelanguage}
Y.~Gao, Y.~Xiong, X.~Gao, K.~Jia, J.~Pan, Y.~Bi et~al.,
  \emph{Retrieval-augmented generation for large language models: A survey},
  2024.

\bibitem{ragreview}
A.J.~Oche, A.G.~Folashade, T.~Ghosal and A.~Biswas, \emph{A systematic review
  of key retrieval-augmented generation (rag) systems: Progress, gaps, and
  future directions},  2025.

\bibitem{rafique2026largelanguagemodelintegration}
A.~Rafique, A.~Singh and R.~Srinivas, \emph{Large language model integration
  for knowledge retrieval and interaction for the dune experiment},  2026.

\bibitem{mallampalli2026mitraaiassistantknowledge}
A.~Mallampalli and S.~Dasu, \emph{Mitra: An ai assistant for knowledge
  retrieval in physics collaborations},  2026.

\bibitem{DalSanto:2935252}
{\scshape ATLAS} collaboration, \emph{{chATLAS: An AI Assistant for the ATLAS
  Collaboration}}, .

\bibitem{Iyer_2024}
K.G.~Iyer, M.~Yunus, C.~O’Neill, C.~Ye, A.~Hyk, K.~McCormick et~al.,
  \emph{pathfinder: A semantic framework for literature review and knowledge
  discovery in astronomy},
  \href{https://doi.org/10.3847/1538-4365/ad7c43}{\emph{The Astrophysical
  Journal Supplement Series} {\bfseries 275} (2024) 38}.

\bibitem{hellert2024physberttextembeddingmodel}
T.~Hellert, J.~Montenegro and A.~Pollastro, \emph{Physbert: A text embedding
  model for physics scientific literature},  2024.

\bibitem{simons2024astrohepbertbidirectionallanguagemodel}
A.~Simons, \emph{Astro-hep-bert: A bidirectional language model for studying
  the meanings of concepts in astrophysics and high energy physics},  2024.

\bibitem{Xu:2025uwt}
X.~Xu, B.~Bolliet, A.~Dimitrov, A.~Laverick, F.~Villaescusa-Navarro, L.~Xu
  et~al., \emph{{Evaluating Retrieval-Augmented Generation Agents for
  Autonomous Scientific Discovery in Astrophysics}},  in \emph{{42nd
  International Conference on Machine Learning}}, 7, 2025
  [\href{https://arxiv.org/abs/2507.07155}{{\ttfamily 2507.07155}}].

\bibitem{Jiang:2026pkh}
R.~Jiang, D.~Fu, C.~Jiang, T.~Yang, Z.~Wang, Y.~Wu et~al., \emph{{Agentic
  Hybrid RAG for Evidence-Grounded Muon Collider Analysis}},
  \href{https://arxiv.org/abs/2606.10381}{{\ttfamily 2606.10381}}.

\bibitem{specter2}
A.~Singh, M.~D'Arcy, A.~Cohan, D.~Downey and S.~Feldman, \emph{Scirepeval: A
  multi-format benchmark for scientific document representations},  2023.

\bibitem{gemma4}
G.~Team, T.~Mesnard, C.~Hardin, R.~Dadashi, S.~Bhupatiraju, S.~Pathak et~al.,
  \emph{Gemma: Open models based on gemini research and technology},  2024.

\bibitem{llama}
H.~Touvron, T.~Lavril, G.~Izacard, X.~Martinet, M.-A.~Lachaux, T.~Lacroix
  et~al., \emph{Llama: Open and efficient foundation language models},  2023.

\bibitem{arxiv_org_submitters_2024}
arXiv.org submitters, \emph{arxiv dataset},  2024.
\newblock 10.34740/KAGGLE/DSV/7548853.

\bibitem{Singh2022SciRepEvalAM}
A.~Singh, M.~D'Arcy, A.~Cohan, D.~Downey and S.~Feldman, \emph{Scirepeval: A
  multi-format benchmark for scientific document representations},  in
  \emph{Conference on Empirical Methods in Natural Language Processing}, 2022,
  \href{https://api.semanticscholar.org/CorpusID:254018137}{https://api.semanticscholar.org/CorpusID:254018137}.

\bibitem{qwen}
Y.~Zhang, M.~Li, D.~Long, X.~Zhang, H.~Lin, B.~Yang et~al., \emph{Qwen3
  embedding: Advancing text embedding and reranking through foundation models},
  {\emph{arXiv preprint arXiv:2506.05176} (2025) }.

\bibitem{nomic}
Z.~Nussbaum, J.X.~Morris, B.~Duderstadt and A.~Mulyar, \emph{Nomic embed:
  Training a reproducible long context text embedder}, {\emph{arXiv preprint
  arXiv:2402.01613} (2024) }.

\bibitem{robertson2009bm25}
S.~Robertson and H.~Zaragoza, \emph{The probabilistic relevance framework: Bm25
  and beyond}, \href{https://doi.org/10.1561/1500000019}{\emph{Foundations and
  Trends in Information Retrieval} {\bfseries 3} (2009) 333}.

\bibitem{cormack2009rrf}
G.V.~Cormack, C.L.A.~Clarke and S.~Buettcher, \emph{Reciprocal rank fusion
  outperforms condorcet and individual rank learning methods},  in
  \emph{Proceedings of the 32nd International ACM SIGIR Conference on Research
  and Development in Information Retrieval}, 2009.

\bibitem{ms-marco-MiniLM-L-6-v2}
N.~Reimers, \emph{Ms marco cross-encoder minilm-l-6-v2},  2021.

\bibitem{nogueira2019passage}
R.~Nogueira and K.~Cho, \emph{Passage re-ranking with {BERT}},  in \emph{arXiv
  preprint arXiv:1901.04085}, 2019.

\bibitem{meta2025llama4}
{Meta AI}, ``The llama 4 herd: The beginning of a new era of natively
  multimodal ai innovation.''
  \url{https://ai.meta.com/blog/llama-4-multimodal-intelligence/}, Apr., 2025.

\bibitem{LeDall:2015ptt}
M.~Le~Dall, M.~Pospelov and A.~Ritz, \emph{{Sensitivity to light weakly-coupled
  new physics at the precision frontier}},
  \href{https://doi.org/10.1103/PhysRevD.92.016010}{\emph{Phys. Rev. D}
  {\bfseries 92} (2015) 016010}
  [\href{https://arxiv.org/abs/1505.01865}{{\ttfamily 1505.01865}}].

\bibitem{Belotsky_2011}
K.~Belotsky, A.~Berkov, A.~Kirillov and S.~Rubin, \emph{Clusters of black holes
  as point-like gamma-ray sources},
  \href{https://doi.org/10.1016/j.astropartphys.2011.05.001}{\emph{Astroparticle
  Physics} {\bfseries 35} (2011) 28–32}.

\bibitem{zheng2023judging}
L.~Zheng, W.-L.~Chiang, Y.~Sheng, S.~Zhuang, Z.~Wu, Y.~Zhuang et~al.,
  \emph{Judging {LLM}-as-a-judge with {MT}-bench and chatbot arena},
  {\emph{Advances in Neural Information Processing Systems} {\bfseries 36}
  (2023) } [\href{https://arxiv.org/abs/2306.05685}{{\ttfamily 2306.05685}}].

\bibitem{Loken_2010}
C.~Loken, D.~Gruner, L.~Groer, R.~Peltier, N.~Bunn, M.~Craig et~al.,
  \emph{Scinet: Lessons learned from building a power-efficient top-20 system
  and data centre},
  \href{https://doi.org/10.1088/1742-6596/256/1/012026}{\emph{Journal of
  Physics: Conference Series} {\bfseries 256} (2010) 012026}.

\end{thebibliography}\endgroup

\end{document}